\documentclass[12pt,preprint]{aastex}

\shorttitle{$g$-mode excitations by SASI}
\shortauthors{Yoshida et al.}

\begin{document}

\title{The $g$-mode Excitation in the Proto Neutron Star by\\ the Standing Accretion Shock Instability}

\author{Shijun Yoshida}  
\affil{Science and Engineering, Waseda University, \\3-4-1 Okubo, Shinjuku, Tokyo 169-8555, Japan}
\email{yoshida@heap.phys.waseda.ac.jp}

\author{Naofumi Ohnishi}

\affil{Department of Aerospace Engineering, Tohoku University,\\
6-6-01 Aramaki-Aza-Aoba, Aoba-ku, Sendai 980-8579, Japan}
\email{ohnishi@cfd.mech.tohoku.ac.jp}

\and 

\author{Shoichi Yamada\altaffilmark{1}}
\affil{Science and Engineering, Waseda University,\\ 3-4-1 Okubo, Shinjuku, Tokyo 169-8555, Japan}
\email{shoichi@waseda.jp}
\altaffiltext{1}{Advanced Research Institute for Science and Engineering,
Waseda University, 3-4-1 Okubo, Shinjuku, Tokyo 169-8555, Japan}

\begin{abstract}
The so-called "acoustic revival mechanism" of core-collapse supernova proposed recently by the Arizona group is an
interesting new possibility. Aiming to understand the elementary processes involved in the mechanism, we have 
calculated the eigen frequencies and eigen functions for the $g$-mode oscillations of a non-rotating proto neutron star. 
The $g$-modes have relatively small eigen frequencies. Since there is a convective region 
inside the proto neutron star, the eigen functions are confined either to the exterior or interior of this evanescent
region although some fraction of eigen functions penetrates it because of their long wave lengths. 
The possible excitation of these modes by the standing accretion shock instability, or SASI, is discussed based on these
eigen functions. We have formulated the forced oscillations of $g$-modes by the external pressure perturbations  
exerted on the proto neutron star surface. The driving pressure fluctuations have been adopted from our previous 
computations of the axisymmetric SASI in the non-linear regime. We have paid particular attention to low $\ell$ modes, 
since these are the modes that are dominant in SASI and that the Arizona group claimed played an important role 
in their acoustic revival scenario. Here $\ell$ is the index of the spherical harmonic functions, $Y_{\ell}^{m}$. 
Although the frequency spectrum of the non-linear SASI is broadened substantially by non-linear couplings, 
the typical frequency is still much smaller than those of $g$-modes, the fact leading to a severe impedance mismatch. 
As a result, the excitations of various $g$-modes are rather inefficient and the energy of the saturated $g$-modes is 
$\sim 10^{50}$erg or smaller, with the $g_2$ mode being the largest in our model. Here the $g_2$ mode has two 
radial nodes and is confined to the interior of the convection region. The energy transfer rate from the $g$-modes 
to out-going sound waves is estimated from the growth of the $g$-modes and found to be $\sim 10^{51}$erg/s 
in the model studied in this paper.
\end{abstract}



\section{Introduction}

The mechanism of core-collapse supernovae has been an open question for decades. In the past 15 years,
a lot of detailed numerical simulations have been done by many authors, paying attention to various aspects of 
the dynamics (see, for example, \citet{kot06} for the latest review of the subject). Spherically symmetric models 
have incorporated full general relativity and solved the radiation-hydrodynamical equations with realistic
equations of state and neutrino physics~\citep{ram00,tho03,lie05,sum05}. These computations demonstrated consistently
that the stalled shock wave is highly unlikely to be revived by neutrino heatings in the spherically symmetric dynamics.

Multi-dimensional simulations have been also making a remarkable progress over the 
times (see \citet{kot06} and references therein).
Local non-spherical motions of convection-type are common in the envelope as inferred from the spectra and 
light curves~\citep{nom06}, and are believed to occur also in the core. The global asymmetry as observed in 
the ejecta of SN1987A is
also supposed to be common to the core-collapse supernovae~\citep{wan02}. Motivated by these observations, 
the implications of the asymmetry for the explosion mechanism have been investigated 
intensively and extensively (see again \citet{kot06} and references therein).

Recently, the instability of the standing accretion shock wave was discovered by \citet{blo03} in the context of 
core-collapse supernova and its nature has been studied both analytically \citep{fog06,yam06} and numerically 
\citep{blo06a,ohn06a,blo06b}. The ramifications to the global asymmetry of the ejecta as mentioned above and 
the high kick velocities of young pulsars~\citep{lyn94} were discussed by \citet{sch04} in more realistic simulations. Since it has been 
demonstrated~\citep{her94,jan96,yam06} over the years that these instabilities tend to facilitate the shock revival 
by neutrino heatings, attempts have been done with little success~\citep{brs06} so far to show by realistic 
multi-dimensional simulations that an explosion is indeed induced.

This situation may change. Very recently, \citet{bur06a,bur06b,bur06c} have proposed a new possibility to induce explosions. 
They have done long-term simulations of axisymmetric 2D radiation-hydrodynamics. Although some microphysics are incomplete and the 
formulation is essentially non-relativistic, their models with no equatorial symmetry and artificial inner boundary condition 
being imposed have obtained a shock revival after a rather long delay ($\gtrsim500$ms after bounce) not by the neutrino heatings 
but by what they call the "acoustic" mechanism. They have claimed that the turbulence of accreted matter induces  
mainly $g$-mode oscillations in the proto neutron star, which in turn emit acoustic waves that are dissipated as they 
propagate outward to become shock waves. 

This is a quite new idea that has eluded the discovery by multi-dimensional numerical studies for many years, the reasons
for which are, the authors think, that previous computations were not long enough and/or put too restrictive conditions 
such as the artificial inner boundary that has been frequently employed to reduce the CFL constraint on numerical 
time steps. \citet{bur06a,bur06b,bur06c} have insisted that the merit of this mechanism is the self-regulation in such a way that 
the $g$-mode excitations and subsequent emissions of acoustic waves are maintained until the shock revival occurs. 
This is an interesting possibility and certainly worth further exploration. 

There are indeed a lot of things remaining to be clarified. First of all, the mechanism seems to be rather inefficient in 
transferring the energy. Inferred from the published results, the explosion energy appears to be substantially smaller than
the canonical value. Since the mechanism constitutes of multiple steps, the efficiency of the couplings between different 
steps should be understood. For that purpose, we think that the linear analysis utilizing an idealized model that mimics the 
realistic situation is more suitable than large scale simulations. In this paper, we calculate the eigen frequencies and eigen functions
of the $g$-modes in a non-rotating proto neutron star. Assuming that the standing accretion shock instability (SASI) is responsible for
the excitation of the $g$-modes in the acoustic mechanism, we estimate the efficiency, formulating and solving the equations for 
forced oscillations of the $g$-modes. In so doing, we use the pressure fluctuations in SASI that we obtained in 
the previous numerical studies~\citep{ohn06a}. The emissions of acoustic waves by these excited $g$-modes are currently 
being investigated numerically and will be published elsewhere~\citep{ohn06b}.

The organization of the paper is as follows. In section~\ref{form}, we describe the models of proto neutron star and 
formulate the basic equations for the normal modes  and their forced oscillations. The SASI that we use as an external
force to drive the oscillations of $g$-modes is also presented there. We give the numerical results in section~\ref{result}.
We conclude the paper in section~\ref{conclusion} with some discussions.

\section{Models \& Formulations\label{form}}

We will investigate the $g$-mode excitations in the proto neutron star, one of the key steps
in the acoustic mechanism, under the following assumptions: (i) A non-rotating proto neutron star 
is approximately in hydrostatic equilibrium. (ii) SASI in the accreting matter is responsible for the 
excitations of $g$-modes. (iii) The oscillation amplitudes in the proto neutron star 
are small compared with the radius. As shown in the numerical simulations done by 
\citet{bur06a, bur06b, bur06c}, these assumptions are reasonable after the stagnation of a prompt 
shock wave until the excited $g$-modes grow sufficiently and their non-linear effects on the dynamics 
become significant. Then the dynamics of the proto neutron star can be treated by the linear perturbation 
theory formulated below.

The linear perturbation of stars can be described in terms of the Lagrangian 
displacement $\mbox{\boldmath{$\xi$}}(t,{\bf r})$, which obeys the equation 
of motion given as
\begin{equation}
{\partial^2\mbox{\boldmath{$\xi$}}\over\partial t^2}+{\bf C}
\cdot\mbox{\boldmath{$\xi$}}={\bf F}\,, 
\label{basic_eq1}
\end{equation}
where ${\bf C}$ and ${\bf F}$ are, respectively, the self-adjoint operator 
and the external force driving the perturbations (see, e.g., Friedman \& Schutz 
1978; Unno et al. 1989). Here, we have assumed the unperturbed star to be static and 
spherically symmetric. Assuming the Lagrangian displacement to be  
\begin{equation}
\mbox{\boldmath{$\xi$}}(t,{\bf r})=\mbox{\boldmath{$\xi$}}_\alpha
({\bf r})e^{i\omega_\alpha t}\,, 
\label{def_xi0}
\end{equation}
where $\mbox{\boldmath{$\xi$}}_\alpha$ and $\omega_\alpha$ are the eigen function and 
eigen frequency, respectively, we can obtain the normal mode solutions, for which the 
eigen function 
and eigen frequency satisfy the following equation,
\begin{equation}
-\omega_\alpha^2\,\mbox{\boldmath{$\xi$}}_\alpha({\bf r})+
{\bf C}\cdot\mbox{\boldmath{$\xi$}}_\alpha({\bf r})=0\,.  
\label{basic_eq2}
\end{equation}
Here, $\{\alpha\}$ denotes the mode index, which includes for spherical stars the angular eigen values 
$\ell$ and $m$ (see below) and the radial mode number $n$. 
We impose for the normal mode solutions the regularity 
condition at the stellar center and $\delta p=0$ at the stellar 
surface, where $\delta p$ stands for the Lagrangian change of the 
pressure $p$. 

When the unperturbed star is static and spherically symmetric, the orthogonality 
of the eigen functions allows us to adopt the normalization 
\begin{equation}
\int \! d^3{\bf r} \, \rho(r) \, \mbox{\boldmath{$\xi$}}^*_{\alpha'}({\bf r})\cdot
\mbox{\boldmath{$\xi$}}_{\alpha}({\bf r})=MR^2 \delta_{\alpha'\alpha} \,, 
\label{norm_eq}
\end{equation}
where $\rho$, $M$ and $R$ denote the density, mass and radius of the unperturbed
star, respectively (see, e.g., Unno et al. 1989). Any Lagrangian displacement 
$\mbox{\boldmath{$\xi$}}$ that satisfies the same boundary conditions as those for the normal 
modes is then expressed as the linear superposition of the normal modes: 
\begin{equation}
\mbox{\boldmath{$\xi$}}(t,{\bf r})=\sum_\alpha A_\alpha(t)
\mbox{\boldmath{$\xi$}}_{\alpha}({\bf r})\,.
\label{expanded_efun}
\end{equation}
Substituting Eq.~(\ref{expanded_efun}) into Eq.~(\ref{basic_eq1}) and 
using Eqs.~(\ref{basic_eq2}) and (\ref{norm_eq}), we obtain the dynamical equation 
for the expansion coefficient $A_\alpha(t)$ as 
\begin{equation}
{d^2 A_\alpha(t)\over dt^2}+\omega_\alpha^2 A_\alpha(t)=
{1\over MR^2} \int \! d^3{\bf r} \,  \rho(r) \, \mbox{\boldmath{$\xi$}}^*_\alpha ({\bf r})
\cdot{\bf F}(t,{\bf r}) \,.
\label{dynamic_eq}
\end{equation}

\subsection{Normal mode oscillations}

In this paper, we study adiabatic oscillations of a non-rotating spherical proto neutron star and 
employ the Cowling approximation, neglecting perturbations of the gravitational potential (see, 
e.g., Unno et al. 1989). For spherical stars, the separation of the angular variables in the 
perturbation equations can be achieved by the use of the spherical harmonics in the spherical polar coordinates  
($r$, $\theta$, $\phi$). The Lagrangian 
displacement vector $\mbox{\boldmath{$\xi$}}$ 
is expressed in terms of the vector spherical harmonics as
\begin{equation}
\left(\xi_r, \, \xi_\theta, \, \xi_\phi\right) = \left(r y_1(r), \, r h(r) {\partial\over\partial\theta}, \, 
r h(r) {1\over\sin\theta}{\partial\over\partial\phi}\right) Y_{\ell}^{m}(\theta,\phi) \, e^{i\omega t}\,, 
\label{def_efun}
\end{equation}
and the Eulerian perturbation of pressure $p'$ is given by 
\begin{equation}
p'=\rho \, g \, r y_2(r) \, Y_{\ell}^{m}(\theta,\phi) \, e^{i\omega t}\,, 
\label{def_ep}
\end{equation}
where $g$ denotes the local gravitational acceleration. Eq.~(\ref{basic_eq2}) is then reduced to 
\begin{equation}
z{d y_1\over dz}=(V_g-3) \, y_1+\left({l(l+1)\over c_1\bar{\omega}^2}-V_g\right)y_2\,,
\label{mseq1}
\end{equation}
\begin{equation}
z{d y_2\over dz}=(c_1\bar{\omega}^2+rA) \, y_1+(-rA-U+1) \, y_2\,,
\label{mseq2}
\end{equation}
where 
\begin{eqnarray}
z&=&r/p\,,\\
V_g&=&{V\over\Gamma}=-{1\over\Gamma}{d\ln p\over d\ln r}\,,\\
U&=&{d\ln M_r\over d\ln r} \,,\\
c_1&=& (r/R)^3/(M_r/M) \,, \\
\bar{\omega}^2&=&\omega^2 R^3/(GM) \,, 
\label{omgbar}\\
rA&=&{d \ln \rho\over d\ln r}-{1\over \Gamma}{d \ln p\over d\ln r} \,. 
\end{eqnarray}
Here, $\Gamma$ and $M_r$ stand for the adiabatic exponent and the stellar mass 
contained inside the radius $r$, respectively.  
The inner and outer boundary conditions are explicitly given by  
\begin{eqnarray}
&&c_1\bar{\omega}^2 y_1-l y_2 = 0 \quad {\rm as} \ r\rightarrow 0\,, 
\label{bc1} \\
&&y_1-y_2=0 \quad {\rm as}\ r\rightarrow R\,. 
\label{bc2}
\end{eqnarray}
The total mode energy $E$ is defined as
\begin{eqnarray}
E={1\over 2}\int \! d^3{\bf r} \, \rho(r) \left(\left|{\partial\mbox{\boldmath{$\xi$}}\over\partial t}\right|^2+
\left|{p'\over\rho c_s}\right|^2+\left({g\over N}\right)^2\left|
{p'\over\Gamma p}-{\rho'\over\rho}\right|^2\right) \,,
\label{emode}
\end{eqnarray}
where $\rho'$ is the Eulerian perturbation of density, and $c_s$ and $N$ mean the sound velocity and 
the Brunt-V\"ais\"al\"a frequency defined, respectively, as 
\begin{eqnarray}
c_s^2=\Gamma p/\rho\,, \quad N^2=-g A\,. 
\end{eqnarray}
In order to obtain eigen frequencies and eigen functions,   
Eqs.~(\ref{mseq1}) and (\ref{mseq2}) are solved numerically as an eigen value
problem for $\bar\omega$, together with the boundary conditions (\ref{bc1}) and (\ref{bc2}).
A Henyey-type relaxation method is employed (see, e.g., Unno et al. 1989).

Introducing new variables $\tilde{\xi}$ and $\tilde{\eta}$ defined as 
\begin{eqnarray}
\tilde{\xi}=r^3 y_1 \exp\left(-\int_0^r{g\over c_s^2} dr \right)\,, 
\end{eqnarray}
\begin{eqnarray}
\tilde{\eta}=gry_2 \exp\left(-\int_0^r{N^2\over g} dr \right)\,,
\end{eqnarray}
we can rewrite the set of the pulsation equations (\ref{mseq1}) and (\ref{mseq2}) 
in a canonical form given by   
\begin{eqnarray}
{d\tilde{\xi}\over dr}=H(r){r^2\over c_s^2}
\left({L_\ell^2\over\omega^2}-1\right)\tilde{\eta}\,, 
\label{caneq1}
\end{eqnarray}
\begin{eqnarray}
{d\tilde{\eta}\over dr}={1\over r^2 H(r)}\,(\omega^2-N^2)\,\tilde{\xi}\,,
\label{caneq2}
\end{eqnarray}
where $L_{\ell}$ is the Lamb frequency defined as 
\begin{equation}
L_{\ell}=c_s {\sqrt{\ell ( \ell +1)}\over r} \,. 
\end{equation}
Here, the function $H(r)$ is positive definite and is defined by
\begin{equation}
H(r)=\exp\left[\int_0^r\left({N^2\over g}-{g\over c_s^2}\right)dr\right]\,. 
\end{equation}

The canonical pulsation equations (\ref{caneq1}) and (\ref{caneq2}) are
more appropriate to gain the insight into the qualitative properties of nonradial oscillations.
Ignoring the spatial variations of all the coefficients in Eqs.~(\ref{caneq1}) and (\ref{caneq2})
and employing a WKB-type local approximation, we obtain the eigen 
functions in the form, 
\begin{equation}
\tilde{\xi}(r),\ \tilde{\eta}(r)\  \propto\  \exp (i k_r r) \,,  
\end{equation}
where $k_r$ means a radial wave number of the perturbations. Then the following 
dispersion relation is obtained:
\begin{equation}
k_r^2={1\over\omega^2 c_s^2}(\omega^2-L_{\ell}^2)(\omega^2-N^2)\,, 
\label{dispersion}
\end{equation}
which implies that if $(\omega^2-L_{\ell}^2)(\omega^2-N^2)>0$, 
the wave number $k_r$ is real and the corresponding wave can propagate in the radial direction, whereas 
the wave number $k_r$ becomes purely imaginary and the amplitude of the perturbation 
grows or damps exponentially with $r$ when $(\omega^2-L_{\ell}^2)(\omega^2-N^2)<0$.

\subsection{Mode excitations}

In this study, we are concerned with the excitation of oscillations in the proto neutron star by 
the turbulence in accreting matter. In our formulation, we take into account the latter effect as pressure
perturbations at the surface of proto neutron star. Hence we set ${\bf F}=0$ in Eq.~(\ref{basic_eq1}) 
and adopt the time-dependent non-vanishing pressure perturbation given by 
\begin{eqnarray}
\delta p = f(t) \, p(r) \, Y_{\ell}^{m}(\theta,\phi) \quad {\rm as} \ r\rightarrow R\,
\label{def_f}
\end{eqnarray}
instead of the ordinary condition $\delta p=0$ at the proto neutron star surface. Here the function of time, $f(t)$, 
describes how the stellar surface is disturbed by the accreted matter and is taken from the numerical results 
obtained in our previous exploration of SASI (see section~\ref{sasi}). 

Note that the boundary condition (\ref{def_f}) is no longer the same as that for the normal modes, $\delta p=0$ at $r=R$.
However, we can still use an expansion by the normal modes, $\mbox{\boldmath{$\xi$}}_\alpha$, for a part of 
the Lagrangian displacement by introducing a new dependent variable $\mbox{\boldmath{$\eta$}}$ defined as 
\begin{eqnarray}
\mbox{\boldmath{$\eta$}}=\mbox{\boldmath{$\xi$}}+\mbox{\boldmath{$\zeta$}} \,, 
\end{eqnarray}
where 
\begin{eqnarray}
\left(\zeta_r,\zeta_\theta,\zeta_\phi\right)=\left(r
\left({1\over\Gamma}\right)_{r=R}{f(t)\over \ell +3}
\left({r\over R}\right)^{\ell} Y_{\ell}^{m}(\theta,\phi), \, 0, \, 0 \right)\,. 
\label{def_zeta}
\end{eqnarray}
Substituting $\mbox{\boldmath{$\xi$}}=\mbox{\boldmath{$\eta$}}-
\mbox{\boldmath{$\zeta$}}$ into Eq.~(\ref{basic_eq1}), we obtain the master equation 
for $\mbox{\boldmath{$\eta$}}$ given by 
\begin{equation}
{\partial^2\mbox{\boldmath{$\eta$}}\over\partial t^2}+{\bf C}\cdot\mbox{\boldmath{$\eta$}}=
{\partial^2\mbox{\boldmath{$\zeta$}}\over\partial t^2}+{\bf C}\cdot\mbox{\boldmath{$\zeta$}}\,. 
\label{basic_eq3}
\end{equation}

In Eq.~(\ref{def_zeta}), $\mbox{\boldmath{$\zeta$}}$ is chosen so that $\mbox{\boldmath{$\eta$}}$ 
could satisfy the same boundary conditions as those for the normal modes: 
\begin{eqnarray}
\delta p_\eta&=&-{p \Gamma\over\rho}\mbox{\boldmath{$\nabla$}}\cdot
\mbox{\boldmath{$\eta$}}=0 \quad {\rm at}\ r=R \,, 
\label{boundary_eq}
\end{eqnarray}
and $\mbox{\boldmath{$\eta$}}$ is regular at $r=0$. As a result, we can express  
$\mbox{\boldmath{$\eta$}}$ as a linear superposition of the eigen functions for the normal modes, 
\begin{equation}
\mbox{\boldmath{$\eta$}}(t,{\bf r})=\sum_\alpha A_\alpha(t)
\mbox{\boldmath{$\xi$}}_{\alpha}({\bf r})\,.
\label{expanded_efun2}
\end{equation}
The expansion coefficient $A_\alpha(t)$ is again determined by the equation similar to 
Eq.~(\ref{dynamic_eq}), 
\begin{equation}
{d^2 A_\alpha\over d\bar{t}^2}+\bar{\omega}_\alpha^2 A_\alpha=
{R\over GM^2}\int \! d^3{\bf r}  \, \rho(r) \, \mbox{\boldmath{$\xi$}}^*_\alpha ({\bf r})\cdot
\left({\partial^2\mbox{\boldmath{$\zeta$}}\over\partial{t}^2}+
{\bf C}\cdot\mbox{\boldmath{$\zeta$}}\right)={R\over GM^2}S_\alpha\,, 
\label{dynamic_eq2}
\end{equation}
where $\bar{t}=t(GM/R^3)^{1/2}$. In Eq.~(\ref{dynamic_eq2}), the driving-force term 
$(R/GM^2)S_\alpha$ is explicitly written as 
\begin{eqnarray}
{R\over GM^2}S_\alpha=af(\bar{t})+b{\partial^2 f(\bar{t})\over\partial\bar{t}^2}\,, 
\label{def_src}
\end{eqnarray}
where $a$ and $b$ are constants determined by the background structure of the proto neutron star 
and the eigen functions for the normal modes and are given as 
\begin{eqnarray}
a&=&\left({1\over\Gamma}\right)_{r=R}\int_0^1\! dx \, {\rho R^3\over M}x^{\ell+2} 
\left[ {Rp\over GM\rho}\left({1\over \ell+3}V_g-1\right)V(y^*_2-y^*_1) \right.
\nonumber \\
&&\quad\quad\quad\quad\quad\quad\quad\quad\quad\quad\quad\quad\quad
\left. -{x^2\over c_1}\left\{{1\over \ell+3}(rA-V_g)+1 \right\}y^*_1 \right]  
\label{def_a} \,, \\
b&=&\left({1\over\Gamma}\right)_{r=R}\int_0^1\! dx \, {\rho R^3\over M}x^{\ell+4}
{y^*_1\over \ell+3} \,. 
\label{def_b}
\end{eqnarray}
Here the normalized radius $x=r/R$ is introduced. 

Once we obtain $A_\alpha(t)$, 
the Lagrangian displacement vector $\mbox{\boldmath{$\xi$}}$ is given by 
\begin{eqnarray}
\mbox{\boldmath{$\xi$}}(t,{\bf r})=
\sum_\alpha A_\alpha(t)\mbox{\boldmath{$\xi$}}_{\alpha}({\bf r})-
\mbox{\boldmath{$\zeta$}}\,. 
\end{eqnarray}
Since $\mbox{\boldmath{$\zeta$}}$ does not satisfy, in general, the orthogonality,
\begin{eqnarray}
\int \! d^3{\bf r}  \, \rho(r) \, \mbox{\boldmath{$\xi$}}^*_\alpha ({\bf r})\cdot 
\mbox{\boldmath{$\zeta$}}=0\,, 
\end{eqnarray}
$\mbox{\boldmath{$\zeta$}}$ includes the components proportional to 
$\mbox{\boldmath{$\xi$}}_\alpha$, and the expansion coefficient 
$A_\alpha(t)$ is only a part of the excited oscillation mode. 
The true amplitude $B_\alpha(t)$ of the excited mode 
$\mbox{\boldmath{$\xi$}}_\alpha$ in $\mbox{\boldmath{$\xi$}}$ can be obtained by
the inner product between $\mbox{\boldmath{$\xi$}}$ and 
$\mbox{\boldmath{$\xi$}}_\alpha$ as 
\begin{eqnarray}
B_\alpha(t)={1\over MR^2}\int \! d^3{\bf r} \, \rho(r) \, \mbox{\boldmath{$\xi$}}^*_\alpha ({\bf r})
\cdot \mbox{\boldmath{$\xi$}}=A_\alpha(t)-
{1\over MR^2}\int \! d^3{\bf r} \,  \rho(r) \, \mbox{\boldmath{$\xi$}}^*_\alpha ({\bf r})\cdot
\mbox{\boldmath{$\zeta$}}\,.
\label{def_Bamp}
\end{eqnarray}
Substituting Eqs.~(\ref{def_efun}) and (\ref{def_zeta}) into Eq.~(\ref{def_Bamp}), we obtain 
\begin{eqnarray}
B_\alpha(t)=A_\alpha(t)-b f(t) \,.
\label{def_Bamp2}
\end{eqnarray}
It is reasonable to assume that no modes are excited in the proto 
neutron star initially, i.e., $B_\alpha(t)=0$ and $dB_\alpha(t)/dt=0$ at $t=0$. Accordingly,  
we start the integration of Eq.~(\ref{dynamic_eq2}) with the following initial conditions,
\begin{eqnarray}
A_\alpha(t)=b f(t) \,, \quad {dA_\alpha(t)\over dt}=b {df(t)\over dt} 
\quad {\rm at} \ \ t=0 \,. 
\label{initial_data}
\end{eqnarray}

\subsection{SASI\label{sasi}}

As already mentioned above, we assume in this paper that oscillations in the proto neutron star 
are excited by the turbulence of accreting matter, which is supposed to be induced by 
SASI. SASI is a non-local hydrodynamical instability that occurs in the 
accretion flow through the standing shock wave onto the proto neutron star~\citep{blo03,sch04,blo06a,ohn06a,blo06b}. 
Although the mechanism is still controversial~\citep{blo06a}, SASI is probably driven by the cycle of inward 
advections of entropy- and vortex-perturbations and outward propagations of pressure fluctuations~\citep{fog06,yam06}.

The prominent feature of SASI is the dominance of low $\ell$ modes both in the linear and non-linear regimes.  
In particular, the $\ell=1$ mode is thought to be a promising cause for the proper motions of young pulsars. In this paper, we regard
SASI as a source of pressure perturbations exerted on the surface of the proto neutron star, that is, $f(t)$ in Eq.~(\ref{def_f}). 

We employ the numerical results obtained by \citet{ohn06a}, in which we have done 2D axisymmetric simulations of the accretion flows
through the standing shock wave onto the proto neutron star. The realistic equation of state by \citet{she98} and 
the standard reaction rates~\citep{bru85} for the neutrino absorptions and emissions on nucleons were employed. The mass 
accretion rate and the neutrino luminosity are the parameters that specify a particular steady accretion flow and were held 
constant during the computations. The self-gravity of the accreting matter was ignored and only the gravitational attraction 
by the proto neutron star was taken into account. The mass of the proto neutron star was also set to be constant (1.4M$_{\odot}$) 
so that we could obtain completely steady background flows, to which velocity perturbations were added initially.

In the following, we use the results for the model with the mass accretion rate of $\dot{M}=$0.19M$_{\odot}$/s and neutrino luminosity of 
$L_{\nu}=3\times10^{52}$erg/s, which are typical for the post bounce core around the shock stagnation. The neutrino spectra 
were assumed to take a Fermi-Dirac distribution with a vanishing chemical potential and a temperature of $T_{\nu_{e}}=4$MeV for
the electron-type neutrino and $T_{\overline{\nu}_{e}}=5$MeV for the electron-type anti-neutrino. The random velocity perturbation 
of less than $1\%$ was added to the radial velocity initially. We refer the readers to \citet{ohn06a} for more detailed formulations.

We observed SASI growing exponentially in the linear phase that lasts for $\sim$100ms and then saturated to enter the non-linear
phase. The $\ell=1$ feature was clearly seen in the non-linear stage for this model although other modes were also generated by 
non-linear mode couplings. We expanded the Eulerian pressure perturbation at the inner boundary by the spherical harmonic functions.
We used only the $\ell=1$ component in this study, since this was claimed to be the most important mode~\citep{bur06a,bur06b,bur06c}.

The power spectrum of this component is shown in Fig.~\ref{fig4}. As can be seen, the pressure perturbation is actually a series of 
impulsive forces that last for quite short times. It should be noted that the linear analysis of SASI for this model gives the eigen frequency
of $\sim30$Hz for the linear $\ell=1$ mode. The non-linear couplings broaden the spectrum both to the lower and higher frequencies.
As seen later in section~\ref{result}, the spectrum has most of the power in much lower frequencies than the typical $g$-mode frequencies
of several hundreds Hz, which leads to a severe impedance mismatch in exciting $g$-modes.

It should be mentioned that the model studied here is not fully self-consistent. For example, what we need in Eq.~(\ref{def_f}) is the
Lagrangian perturbation of pressure while what we obtained above is the Eulerian perturbation. This is simply because the former
is difficult to obtain from the numerical data. Although they are different, rigorously speaking, we think that the difference will not 
change the outcome qualitatively, since it is the typical frequency and spectral feature of SASI that matters in this study. We should 
also bear in mind that the SASI spectrum employed in this study is just a single realization of essentially chaotic flows 
although we expect it is representative.

\subsection{Proto neutron star model}

Finally, the spherically symmetric, stationary proto neutron star that is adopted as the unperturbed state in this study 
is briefly described. The model is based essentially on the results of detailed radiation-hydrodynamical simulation of 
core-collapse by \citet{sum05}, but we have modified it in the following way.

We have developed a numerical code to compute the long-term post-bounce evolution in the quasi-steady approximation~\citep{wat06},
the details of which will be published elsewhere. In this formalism, we solve separately but consistently three different regions
in the collapsing star simultaneously: (1) the proto neutron star that cools in a quasi-static way. We calculate both the equilibrium 
configuration and the neutrino transport in it with matter accretion taken into account. (2) the steady accretion flow from the standing 
shock wave down to the proto neutron star for the given neutrino luminosity obtained in the calculation (1) and matter accretion rate
given by the next computation. (3) the dynamical implosion of the envelope toward the standing shock wave. The second and third regions are 
joined by the Rankine-Hugoniot condition at the shock wave consistently. 

The code can follow the spherically symmetric quasi-static evolutions for more than 10 seconds and we have confirmed that 
the evolution is well reproduced for the first second, for which the result of the detailed dynamical simulations is available. In this study, 
we utilize the proto neutron star, which is obtained in this approximation and is in hydro-static equilibrium at $\sim$360ms after bounce. 
Physical quantities characterizing the proto neutron star model, that is,  mass $M$, radius $R$, and Kepler frequency $\Omega_{\rm k}$ 
are given in Table~\ref{table1}. The profiles of density, pressure, specific entropy, and 
electron fraction are shown in Fig.~\ref{fig0} as a function of radius. It is noted that 
this spherically symmetric model contains a convectively unstable region around $r\sim20$km.

Here again, the proto neutron star model is not very consistent with the model for SASI. As mentioned in the previous section, 
it is not our aim to construct a fully self-consistent model in this paper, where the elementary feature of the $g$-mode excitations is
explored qualitatively from an experimental point of view. For that purpose, the slight inconsistency in the current model does not 
matter.

\section{Results\label{result}}

We first consider the oscillation properties of $g$-modes for the proto neutron 
star model described in the previous section. Before giving the eigen functions, we look into the so-called propagation diagram 
(see, e.g., Unno et al. 1989), which is useful to see where in the star the wave with a particular frequency $\omega$ can propagate. The 
diagram shows the square of the Lamb frequency $L_{\ell}^2$ and the Brunt-V\"ais\"al\"a frequency $N^2$ as a function of the normalized 
radius $r/R$. 
As explained in section \ref{form}, since the waves can propagate when $(\omega^2-L_{\ell}^2)(\omega^2-N^2)>0$ is satisfied, 
the region where this condition holds true is called the propagation zone, and the other regions in the diagram are referred to 
as an evanescent zone, in which the perturbations do not oscillate but grow or damp exponentially.
 
Fig.~\ref{fig1} is the propagation diagram for the present proto neutron star model, in which we take 
$\ell=1$ to calculate the Lamb frequency and the frequency is normalized according to Eq.~(\ref{omgbar}). 
The propagation regions can be classified into two categories: 
one is the $g$-mode propagation zone indicated as hatched regions and the other is 
the $p$-mode propagation zone shown as a cross-hatched region in Fig.~\ref{fig1}. 
As mentioned already, since the proto neutron star has a convective region around $r\sim 20 {\rm km}$, 
which is an evanescent zone for $g$-modes, there are two disconnected $g$-mode propagation zones, one is near the center of 
the star and the other near the surface. This implies that a particular $g$-mode is almost trapped in one of these  
two disconnected $g$-mode propagation zones. In other words, the $g$-modes for the current proto neutron star model are 
essentially divided into two classes, i.e., what we refer to as the core $g$-modes ($g^c$-modes) and the surface $g$-modes 
($g^s$-modes), whose amplitudes of the eigen functions have a peak near the center and the surface of the proto neutron star, 
respectively.

We now discuss the properties of the normal mode oscillations obtained numerically for the current proto 
neutron star model. In Table~\ref{table2}, we summarize the key quantities for several low-order $g$-modes: 
the normalized angular eigen frequencies, $\bar{\omega}$, the corresponding frequencies in Hz, $\nu$,  
the mode energies, $E$, and the integral quantities, $a$ and $b$, defined by Eqs.~(\ref{def_a}) and (\ref{def_b}).  
The last two are important in evaluating the excitations of $g$-modes by external perturbations in the next section. 
The mode energy gives us a rough measure of non-linear oscillations in the proto neutron star. Note that 
we employ the normalization condition for the eigen functions, Eq.~(\ref{norm_eq}), in evaluating the mode energy $E$.  
As expected, the mode energy is of the order of $10^{53}$ergs. The typical frequency of low-order 
$g$-modes is several hundreds Hz, and we can confirm that the numbers given in Table~\ref{table2}  are comparable 
with those found by \citet{bur06a}. It should be also emphasized that these frequencies are much 
larger than the typical frequency, $\lesssim$100 Hz, of SASI in the non-linear stage. 
It is also recognized in Table~\ref{table2} that for $\ell = 1, 2$ the lowest mode is a surface mode and core
modes and surface modes appear alternately as one goes to higher overtones. 

In Fig.~\ref{fig2}, the eigen functions, $y_{1}$ and $y_{2}$ in Eqs.~(\ref{def_efun}) and (\ref{def_ep}), are shown for 
the $g$-modes with $\ell=1$ as a function of the normalized radius 
of the proto neutron star $r/R$. In the figure, the $g_n$-mode means the $n$-th radial overtone 
$g$-mode, which is characterized by $n$ radial nodes. As mentioned just above, the $g_1$- and $g_2$-modes are 
a typical surface and core $g$-mode, respectively. In fact, the eigen functions have a non-negligible amplitude either 
outside or inside the convection region, which is an evanescent zone for $g$-modes and located at $r \sim 20$km. 
This is also true for other $g$-modes with different $\ell$ or $n$. 
Upon using these eigen functions, the quantities, $a$ and $b$, which appear in Eq.~(\ref{def_src}) and give the driving force 
for mode excitations, are obtained by integrating Eqs.~(\ref{def_a}) and (\ref{def_b}). Although the general trend is that 
their absolute values decrease as the radial node number increases, there are exceptions that are critically important 
in determining which mode is most efficiently excited by external perturbations as explained below.

\begin{figure}
\epsscale{.80}
\plotone{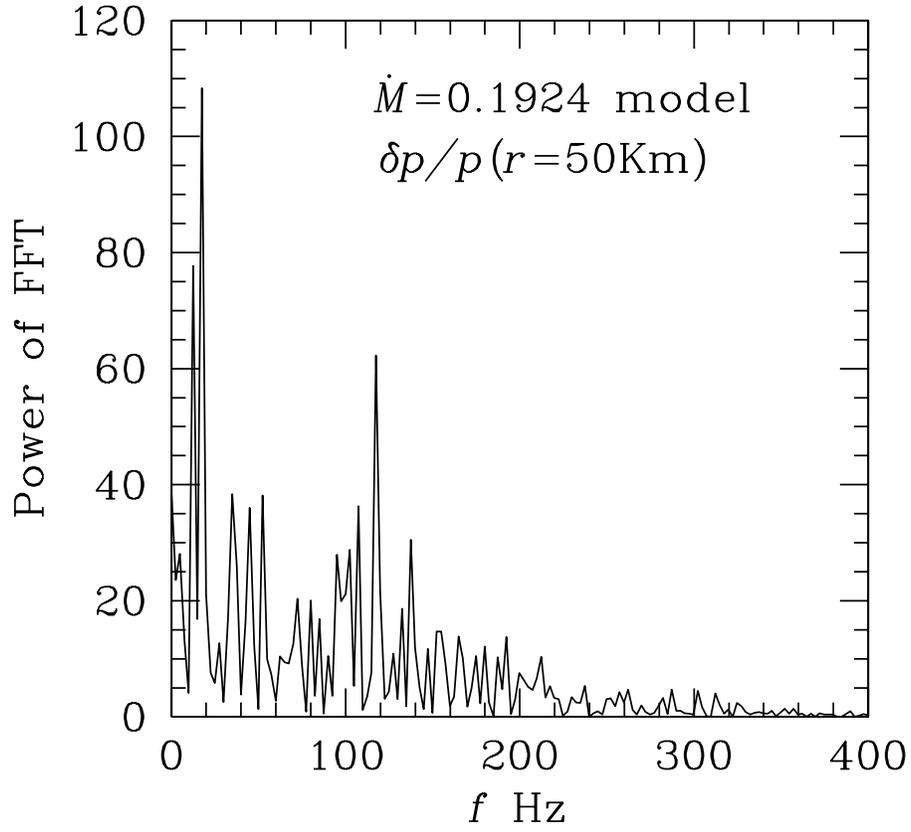}
\caption{Power spectrum of the fractional pressure variation at $r=50 {\rm km}$, 
given as a function of the frequency. Here, the fractional pressure variation is 
expanded in terms of spherical harmonics, and the fractional pressure variation 
associated with $\ell=1$ and $m=0$ is displayed. \label{fig4}}
\end{figure}

\begin{figure}
\epsscale{1.10}
\plottwo{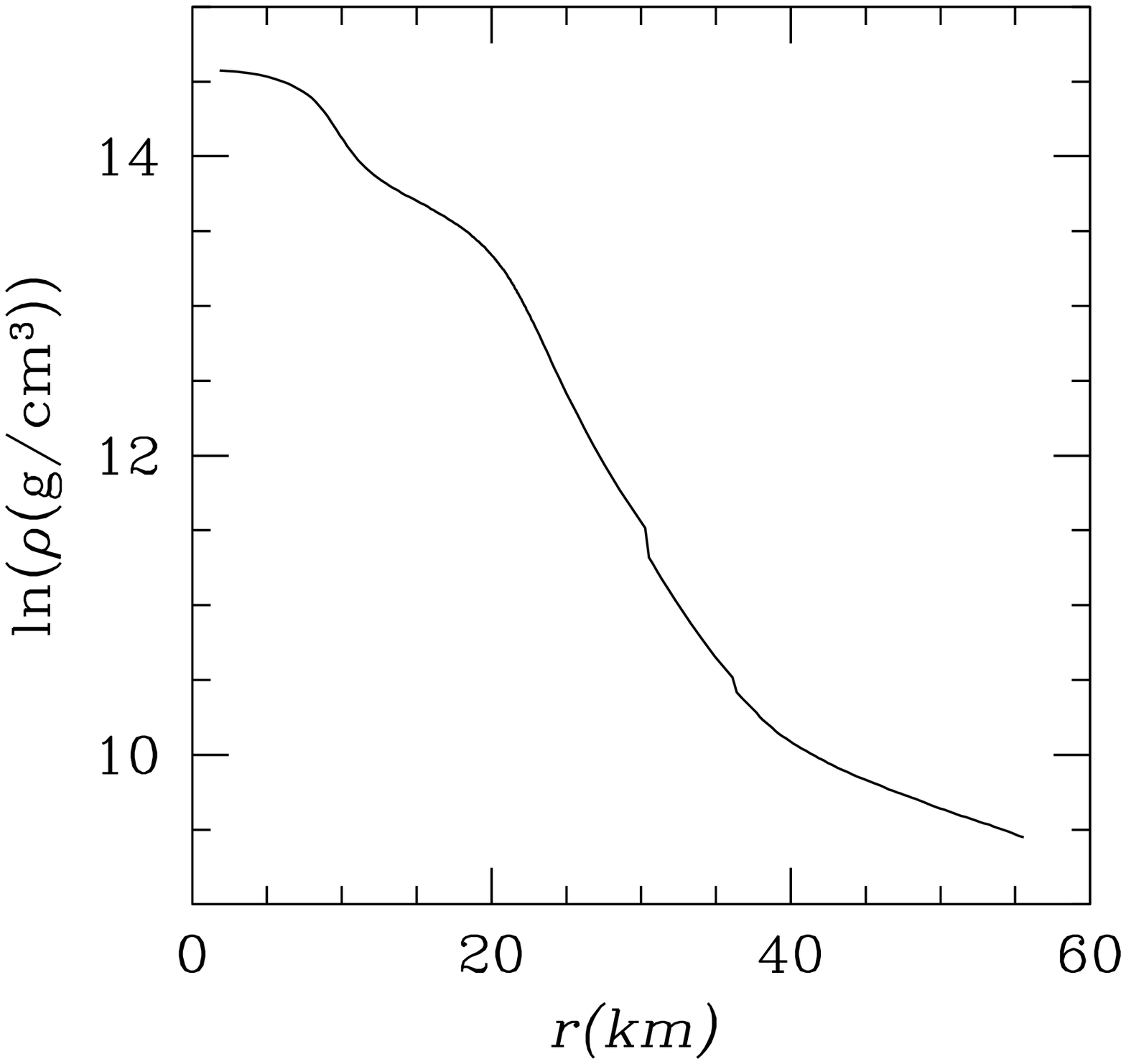}{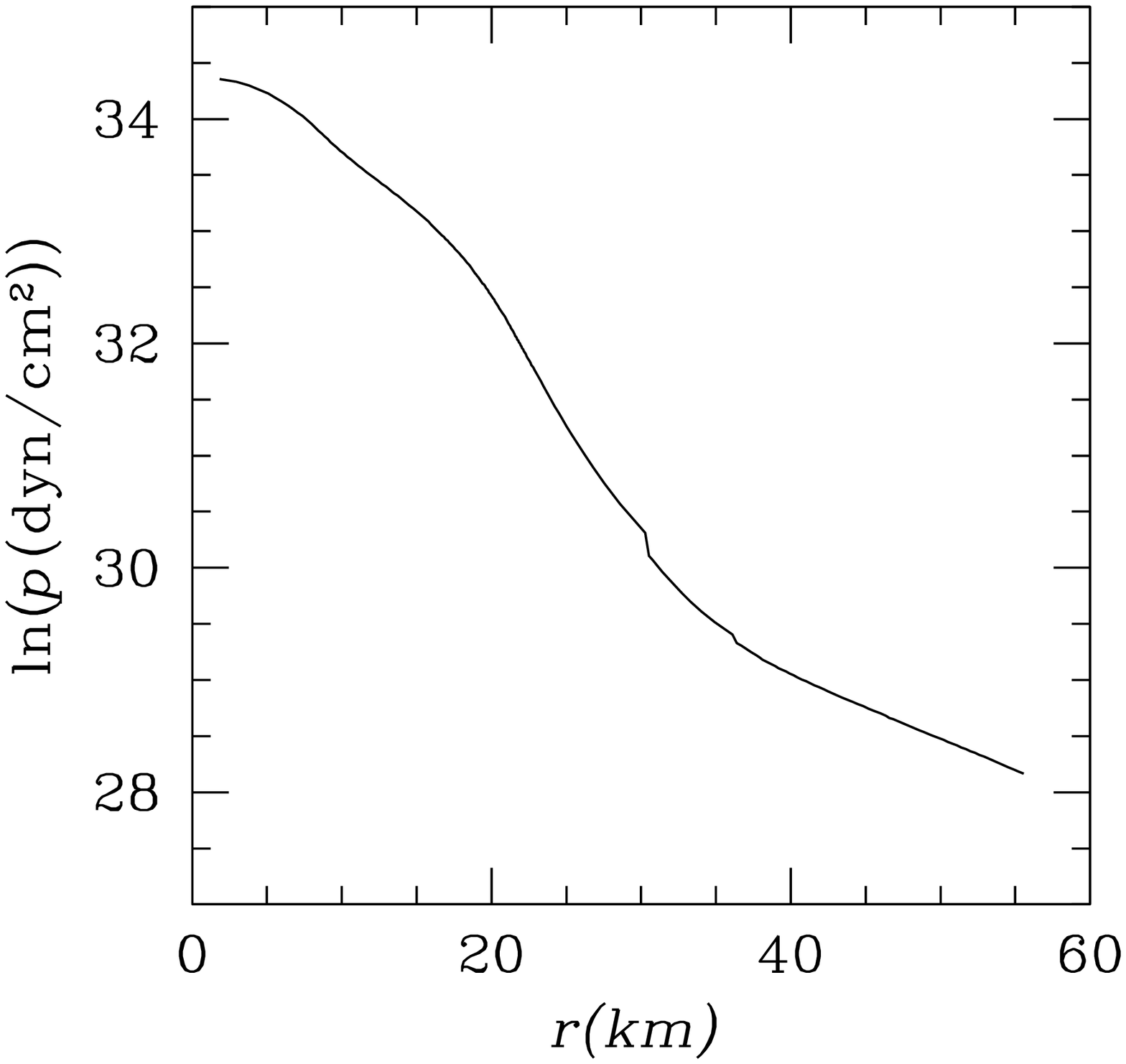}
\plottwo{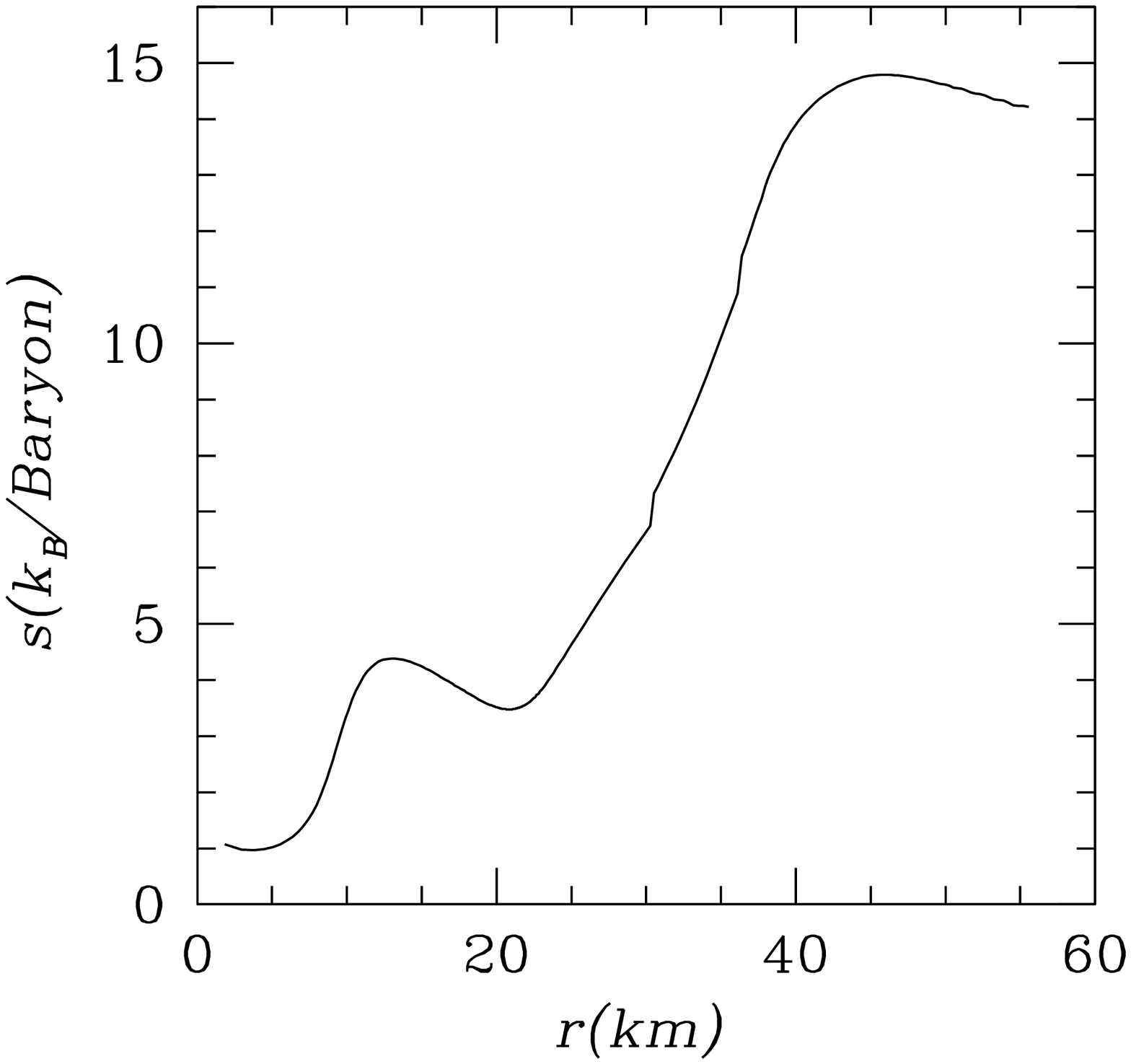}{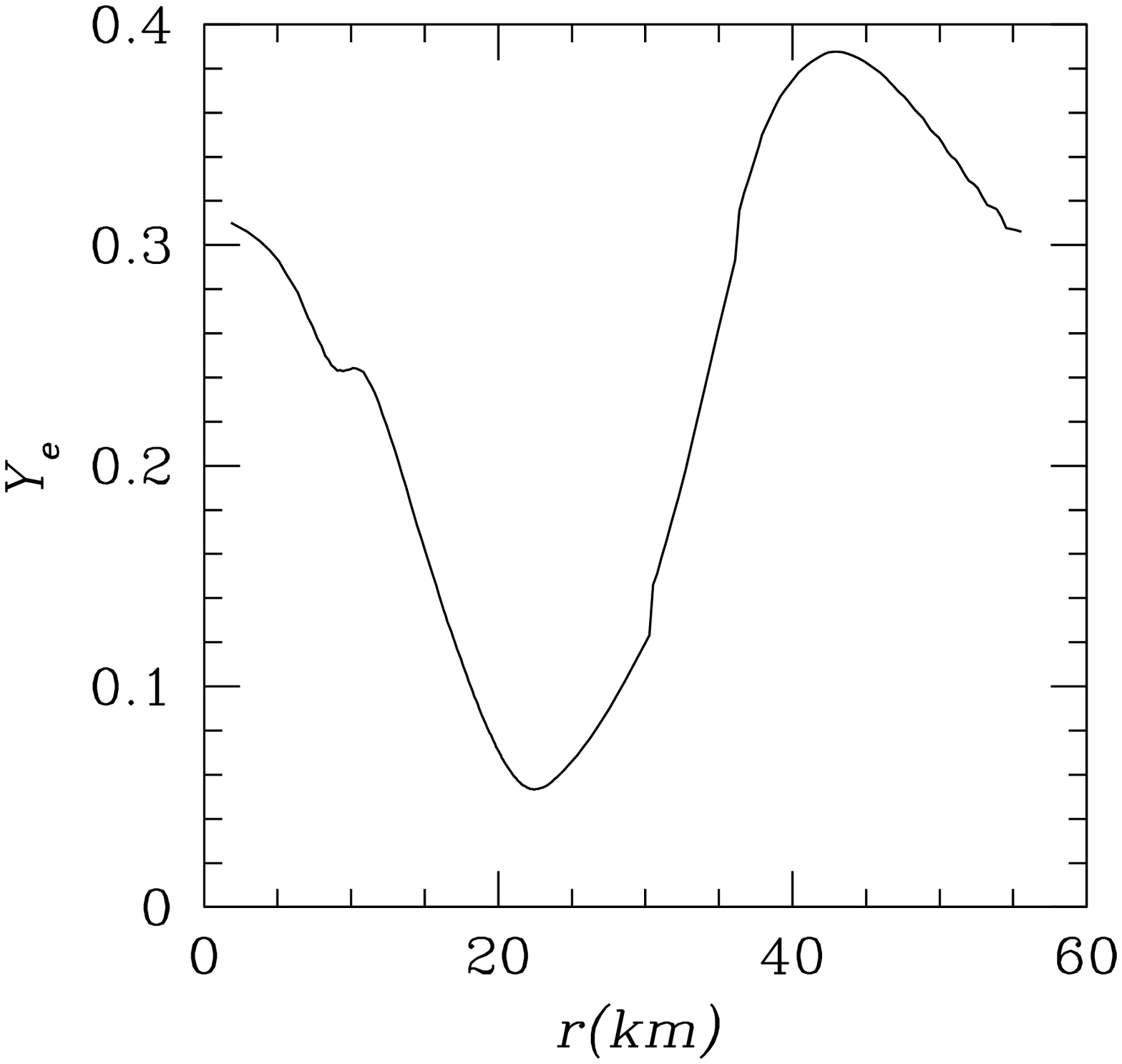}
\caption{Profiles of density $\rho$ (upper left panel), pressure $p$ (upper right panel), 
specific entropy $s$ (lower left panel), and electron fraction $Y_e$ (lower right panel)
for the proto neutron star model considered in this study. Here, $k_B$ denotes 
the Boltzmann constant. \label{fig0}}
\end{figure}

\begin{figure}
\epsscale{.80}
\plotone{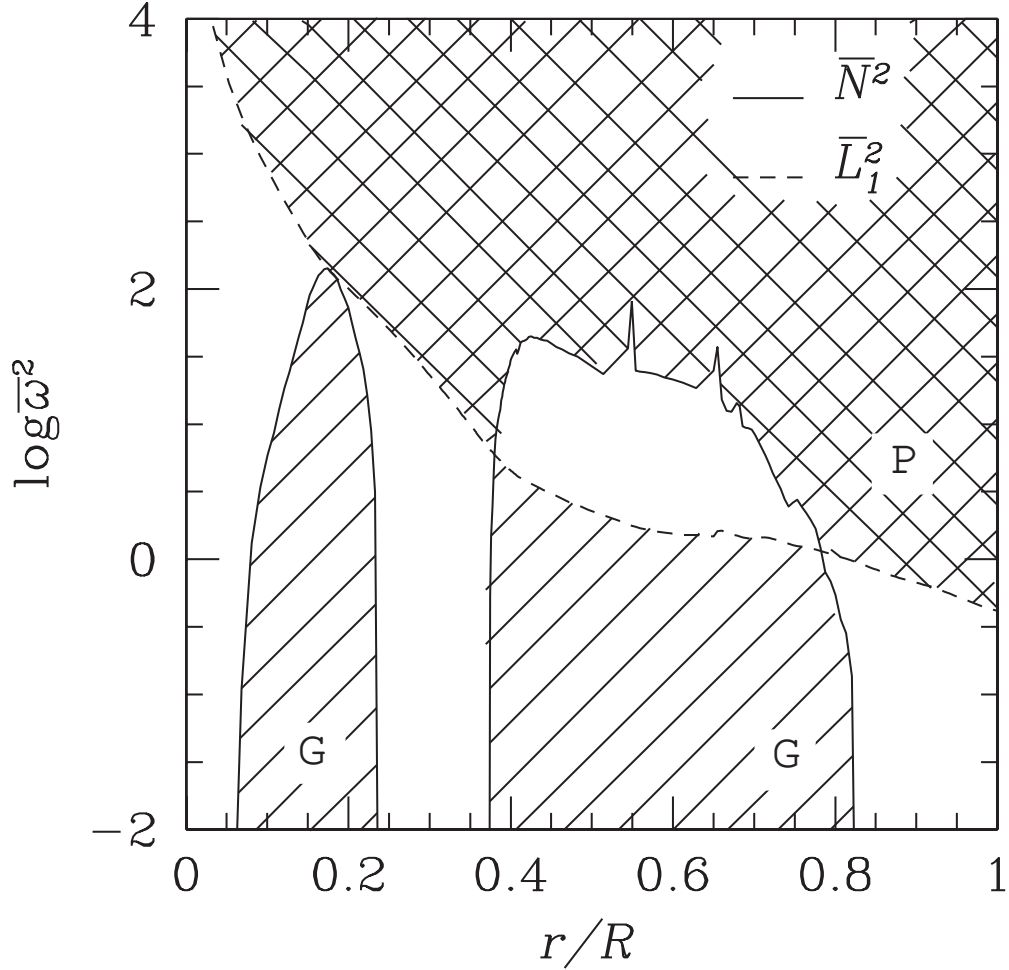}
\caption{Propagation diagram for the proto neutron star model given in Fig.~\ref{fig0}, in which the 
square of the Lamb frequency $L_{\ell}^2$ for $\ell=1$ and the Brunt-V\"ais\"al\"a frequency $N^2$ are shown 
as a function of the normalized radius of the star $r/R$. The frequency is normalized according to Eq.~(\ref{omgbar}).
The $g$- and $p$-mode propagation zones are indicated as the hatched and cross-hatched regions, respectively. 
\label{fig1}}
\end{figure}

\begin{figure}
\epsscale{1.10}
\plottwo{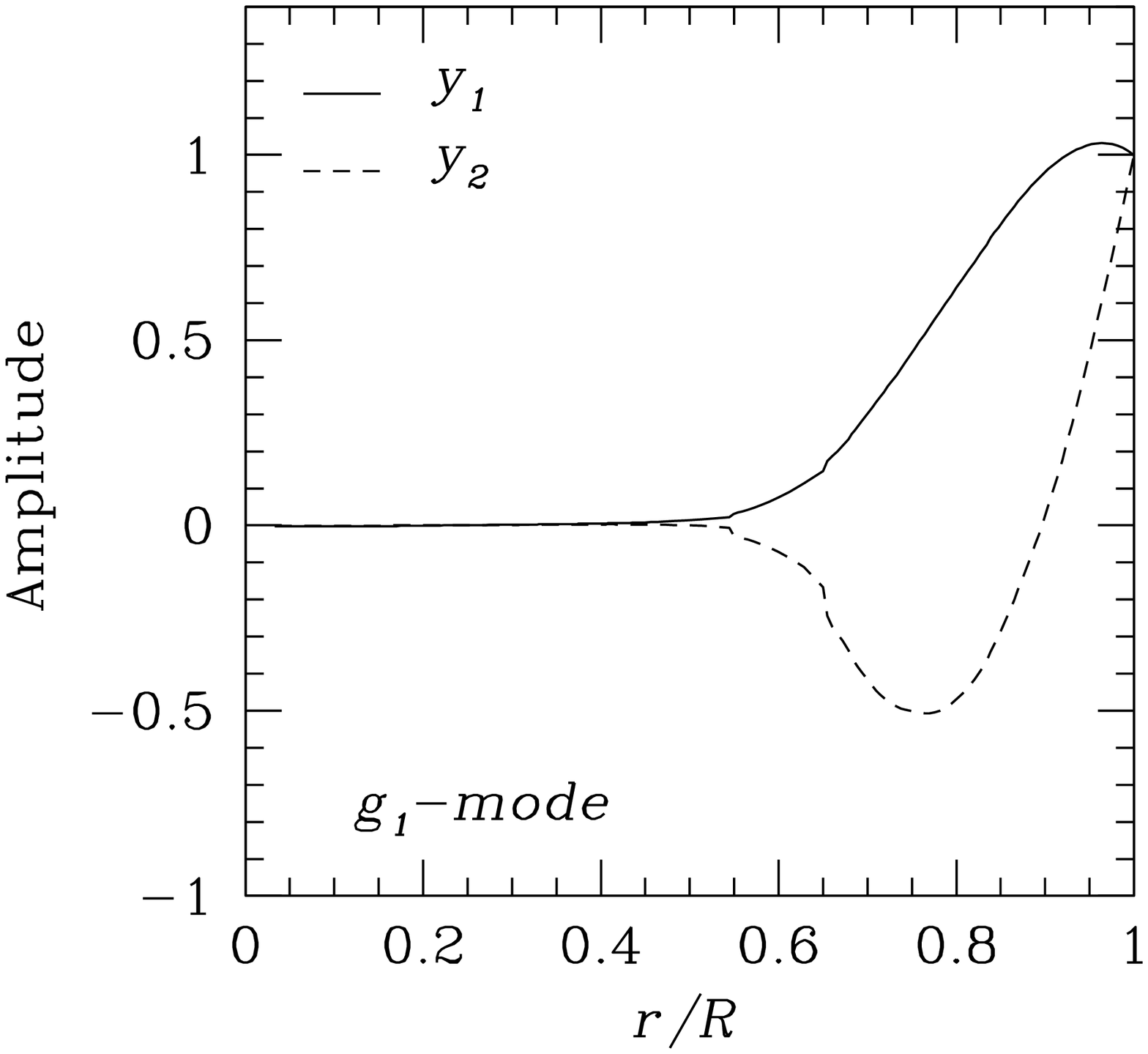}{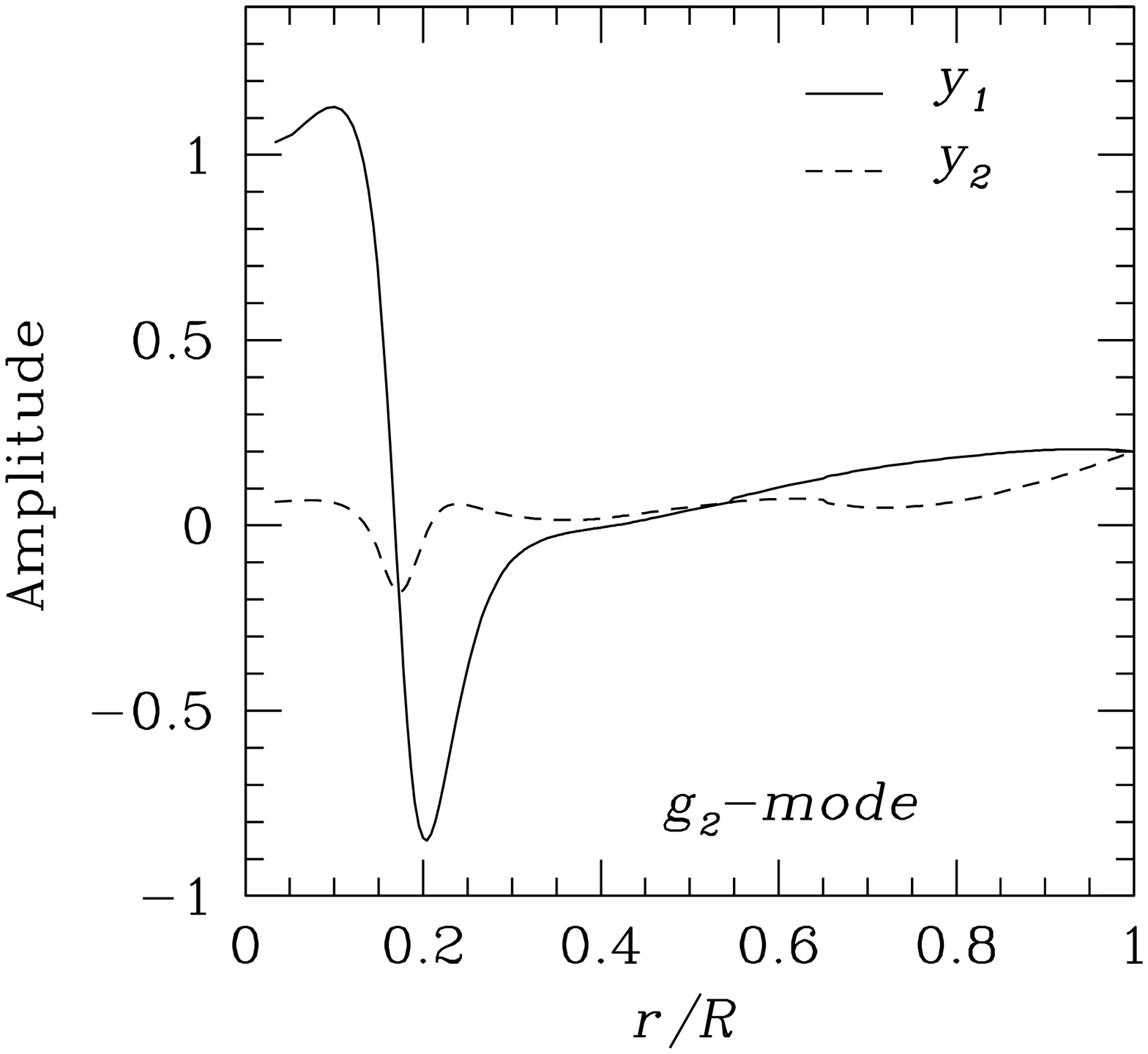}
\caption{Eigen functions $y_1$ and $y_2$ as a function of the normalized radius $r/R$ for the $g_1$-mode (left) and 
$g_2$-mode (right) with $\ell=1$ for the proto neutron star model given in Fig.~\ref{fig0}.\label{fig2}}
\end{figure}

Now we move on to the excitation by accreting matter of the $g$-modes considered just above. 
As mentioned before, we assume in this study that the $g$-modes are excited by the pressure fluctuations around 
the surface of the proto neutron star, which are induced by SASI in the accretion flow. The pressure fluctuation $f(t)$ 
in Eq.~(\ref{def_f}) is obtained from the numerical results on SASI and is displayed in Fig.~\ref{fig4} as a power spectrum. 
It should be repeated that the major peaks of the spectrum are observed in the frequencies much lower than those of 
the low-order $g$-modes given in Table~\ref{table2}. In other words, the typical frequency of SASI, the driving force, 
is much lower than that of the low-order $g$-modes that should be excited in the proto neutron star. It is true that 
these numbers are dependent on the models of proto neutron star and SASI as well as EOS employed, but we expect that the 
mismatch of the frequencies is a rather generic issue. 

By making use of the pressure fluctuations shown in Fig.~\ref{fig4} and solving Eq.~(\ref{dynamic_eq2}) together with 
the initial condition given in Eq.~(\ref{initial_data}), we calculate the time evolution of 
the mode amplitude, $B_\alpha (t)$, for several $g$-modes. Among them, the most efficiently excited $g$-mode is 
the $g^c_2$-mode with $\ell=1$. This is mainly because it has the largest $a$ as given in Table~\ref{table2}. 
In Fig.~\ref{fig3}, we show the temporal evolution of the mode energy for the $g^c_2$-mode with $\ell=1$. 
Since the driving forces, that is, the pressure fluctuations induced by SASI are a collection of impulsive forces, the mode energy is 
fluctuating on a short time scale. It is clear, however, that the mode is growing for $\sim$100ms initially and then is saturated. 
The maximum value of the excited mode energy is $\lesssim 10^{50} ({\rm erg})$ for this mode, which is considerably smaller than 
the typical energy of non-linear oscillations, $\sim10^{53}$erg, given in Table~\ref{table2}. In this sense, the excitation of the $g$-modes
is inefficient, at least for the present model. This is a direct consequence of the severe mismatch in the frequencies of the driving force, SASI, and
the excited $g$-modes. Hence it is more appropriate to refer to what we see here as the forced oscillation rather than as the excitation. 
It is noted, however, that the Eulerian pressure perturbations at $r\sim50$km are found to be of the same order as those found in 
\citet{bur06a}.

The maximum values of the mode energy for other $g$-modes, $\lesssim 10^{49}$erg, are even smaller than 
that of the $g^c_2$-mode with $\ell=1$ for the proto neutron star model considered here. We show in Fig.~\ref{fig6} the temporal evolutions of the
mode energy for the $g_1$- and $g_2$-modes with $\ell=1$. The essential feature is the same as for the $g_2$-mode
shown in Fig.~\ref{fig3}. These modes also grow for about 100ms initially and then get saturated with large fluctuations on a short time scale.
It is emphasized again that all these excitations are non-resonant. In that case, the maximum amplitude of $\partial^{2}\!A_\alpha(t)/\partial t^{2}$ can 
be approximated by the maximum value of $a f(t)$, which, for example, becomes $\sim 10^{-2}$ for the $g_2$-mode and an-order-of-magnitude 
smaller for other modes with  $\ell=1$. From Eq.~(\ref{emode}), we then have approximate values of the mode energies, which are consistent with 
the numerical results given in Figs.~\ref{fig3} and \ref{fig6}.

\section{Summary and Discussions \label{conclusion}}

Inspired by a series of papers on the acoustic mechanism of core-collapse supernova by \citet{bur06a,bur06b,bur06c},
we have calculated in this paper the eigen frequencies and eigen functions for several low-order $g$-modes in the proto neutron star, 
which were claimed to play an important role in the acoustic mechanism, 
and have studied also their excitations by SASI in the accreting matter. We have constructed a proto neutron star model, 
which is supposed to represent a quasi-equilibrium configuration around a few hundred milliseconds after bounce, based on the detailed
radiation-hydrodynamical simulations by \citet{sum05}. We have employed the temporal variation of pressure on surface of 
the proto neutron star, which were obtained in the study on SASI by \citet{ohn06a}.

We have found that the $g$-modes are classified either as the surface $g$-mode or as the core $g$-mode according to the profile of the 
eigen function, since there is a convective zone in the proto neutron star, which is an evanescent zone for the $g$-modes.
The eigen frequencies of the low-order $g$-modes are several hundred Hz, which are larger in general than the typical frequency of 
SASI, $\lesssim$100Hz, in the non-linear regime. 

Assuming that SASI is the driving force for the $g$-mode excitations, we have solved numerically the equations for the time evolution 
of the $g$-mode amplitude with the pressure perturbations at the surface of the proto neutron star taken into account. 
We have observed that the $g$-mode amplitudes grow for $\sim$100ms initially and then get saturated. For the proto neutron star model considered in this paper, 
the $g^c_2$-mode, that is, the second overtone core $g$-mode with $\ell=1$ is found to have the largest mode energy at saturation.
Even for this mode, the mode energy, $\lesssim10^{50}$erg, is much smaller than the typical energy, $\sim10^{53}$erg, 
for non-linear oscillations, which can be estimated by imposing the natural normalization condition to the eigen functions.
This apparent inefficiency of the $g$-mode excitations is a direct consequence of the mismatch of the frequencies of 
the $g$-modes and SASI. 

\begin{figure}
\epsscale{.80}
\plotone{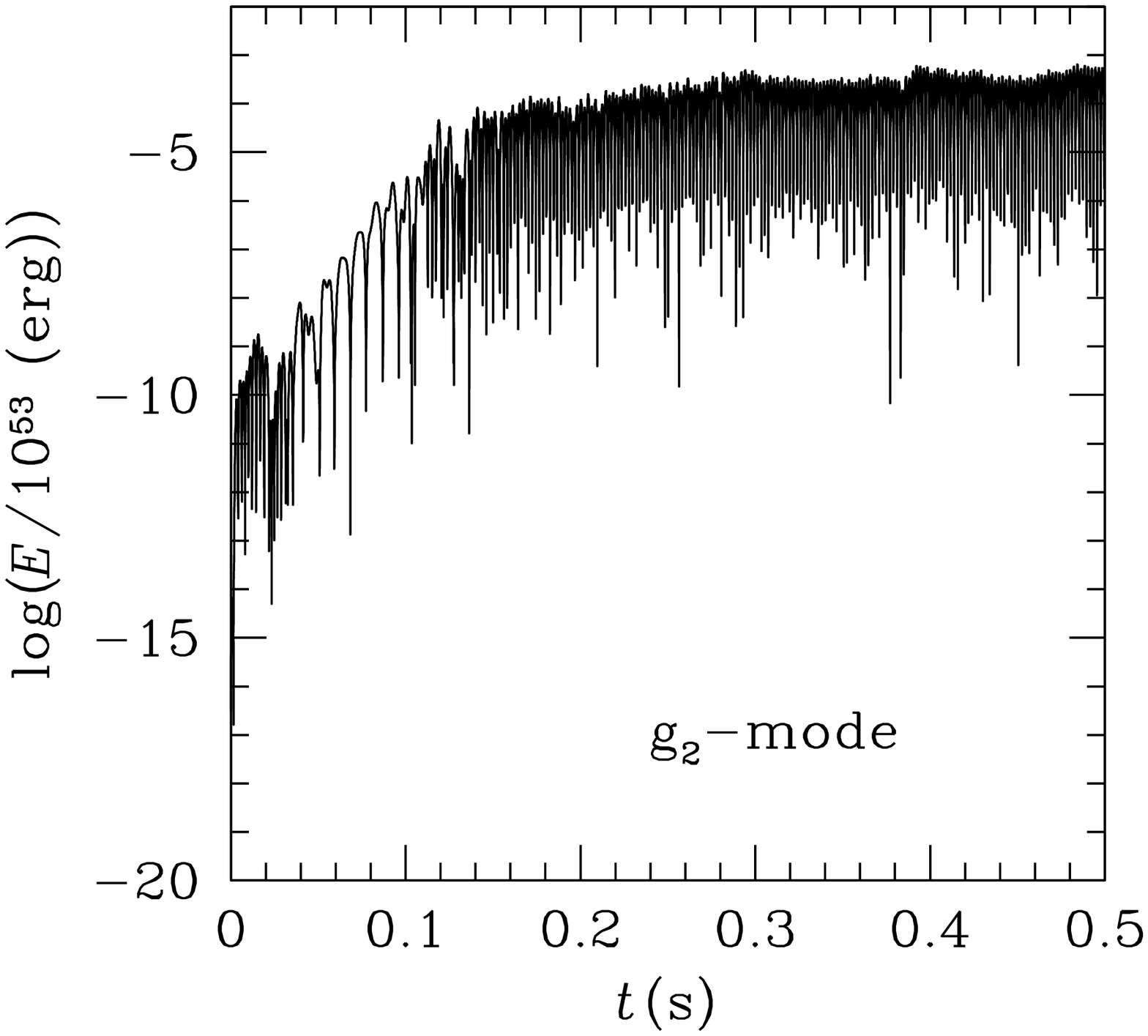}
\caption{Time evolution of the mode energy for the $g_2$-mode with $\ell=1$ for the
proto neutron star model given in Fig.~\ref{fig0}. \label{fig3}}
\end{figure}

\begin{figure}
\epsscale{1.10}
\plottwo{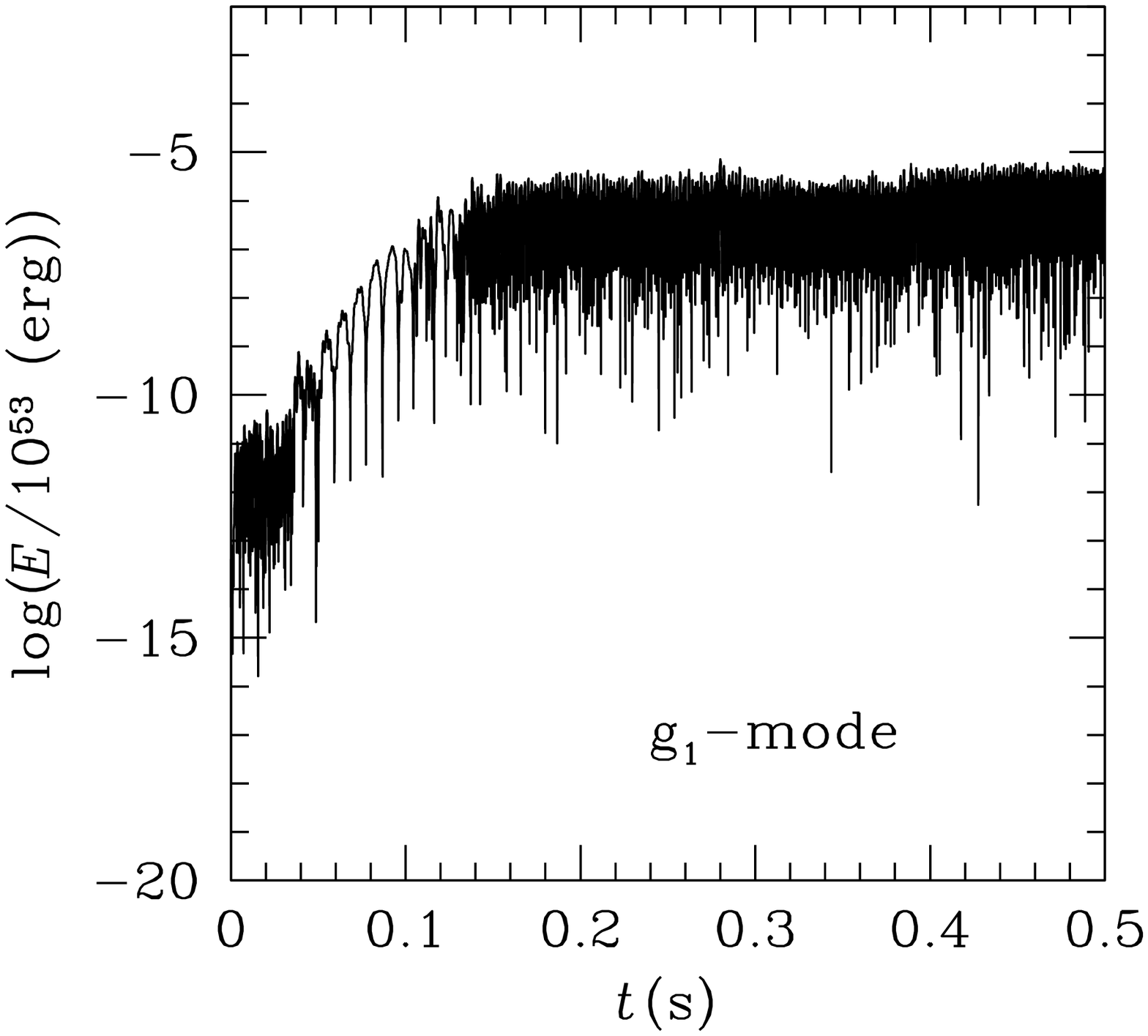}{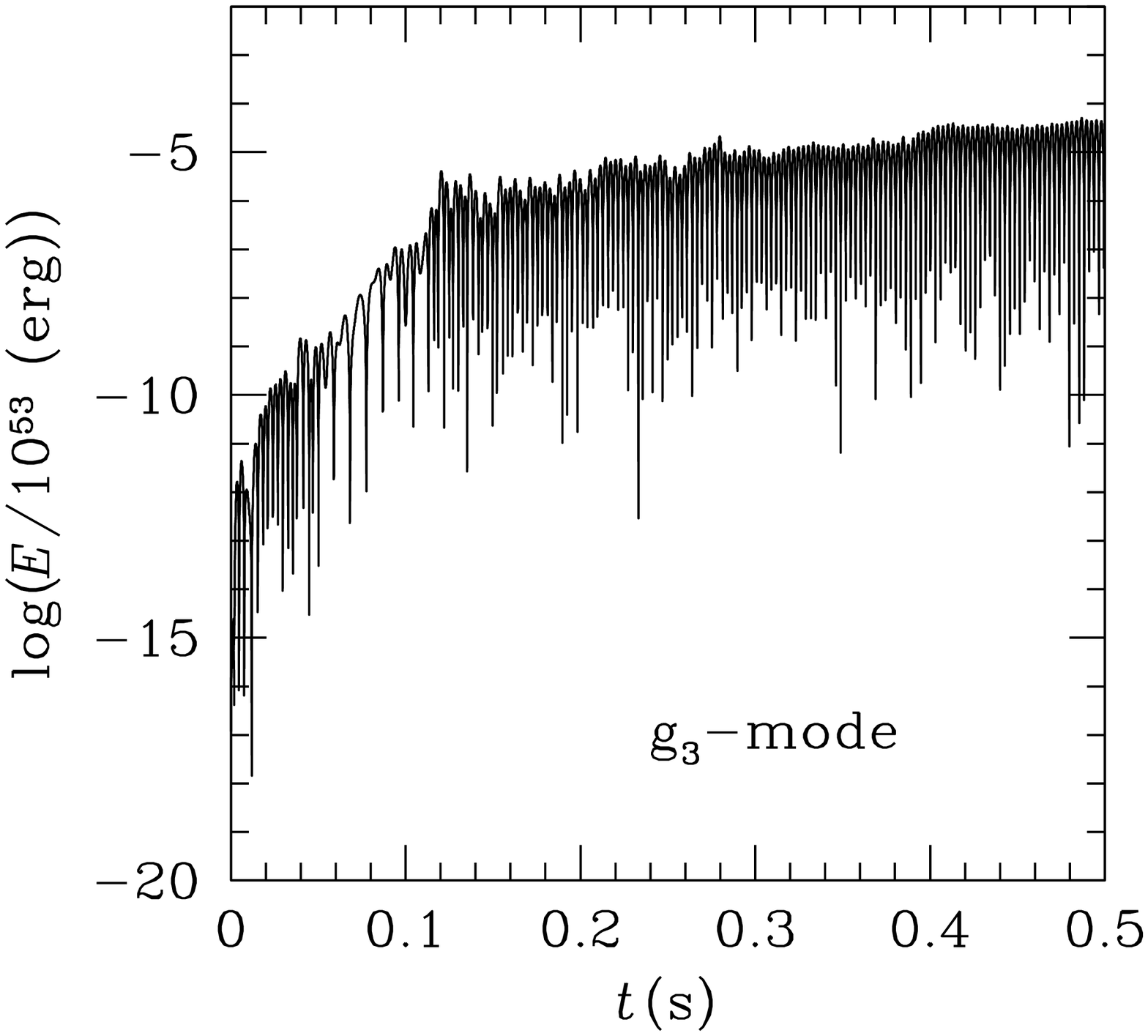}
\caption{Time evolution of the mode energy for the $g_1$-mode (left) and $g_3$-mode 
(right) with $\ell=1$ for the
proto neutron star model given in Fig.~\ref{fig0}. \label{fig6}}
\end{figure}

It is noted, however, that what is important for the acoustic mechanism may not be the mode energy. In fact, the \citet{bur06a} 
mentioned that the $g$-mode oscillations acted like a transducer that transformed the gravitational energy of accreted matter to 
the energy of out-going sound waves. In the present formulation, unfortunately, it is in principle impossible to obtain the energy 
that will be transferred to the out-going acoustic waves. However, we may be able to infer it as follows. In the growth phase that lasts
for $\sim$100ms, the energy of accreting matter, or the work done by the external pressure perturbation in the present model, is 
transferred to the mode energy, as is apparent in Fig.~\ref{fig3}. After saturation, on the other hand, the mode energy is almost unchanged.
In this phase, then, it may be that the energy of the accreting matter is emitted just at the same rate as the accumulation rate. 
If this is really the case, we can estimate that the emitted energy, possibly as out-going acoustic waves, will be $\sim10^{51}$erg/s,
which may not be a bad estimate, provided, in general, the explosion is considerably delayed ($\gtrsim$500ms) in the acoustic mechanism.
It is obvious, however, that more elaborate formulation of the problem, which would include the out-going waves, is required. 
Numerical approaches from an experimental view point will be also useful~\citep{ohn06b}.

There are many other improvements remaining to be done. As repeated in the text, we have not attempted to make the model 
fully self-consistent in this paper, since the qualitative understanding of the $g$-mode excitations by SASI is the focus of the
paper. However, the dependence of the results on the proto neutron star model, EOS, the progenitor model that determines the 
accretion rate, and so forth should be investigated in addition to making the model more self-consistent, all of which are currently 
under way.

\acknowledgments

This work is partially supported by the Grant-in-Aid for the 21st century
COE program "Holistic Research and Education Center for Physics of
Self-organizing Systems" of Waseda University and for Scientific Research
(14740166, 14079202, Young Scientist (B) 17740155) of the Ministry of Education, 
Science, Sports and Culture of Japan.



\clearpage

\begin{deluxetable}{cccc}
\tabletypesize{\scriptsize}
\tablecaption{The mass, radius and Kepler frequency for the proto neutron star model.\label{table1}}
\tablewidth{0pt}
\tablehead{  & $M/M_\odot$ & $R ({\rm km})$ & $\Omega_{\rm k} ({\rm Hz})$ }
\startdata
  & 1.492 & 55.6 & 1074 \\
\enddata
\end{deluxetable}

\begin{deluxetable}{ccccccrr}
\tabletypesize{\scriptsize}
\tablecaption{Key quantities for $g$-modes in the proto 
neutron star.\label{table2}}
\tablewidth{0pt}
\tablehead{ & $\ell$ & Mode & $\bar{\omega}$ & $\nu$ (Hz) & $E (10^{53}{\rm erg})$ & $a$ & $b$}
\startdata
 & 1 & $g^{\rm s}_1$ & 2.92 & 499 & 5.70 & 3.50E-03 &    1.72E-03 \\
 &   & $g^{\rm c}_2$ & 1.43 & 245 & 1.30 & 2.13E-02 & $-$2.99E-04 \\
 &   & $g^{\rm s}_3$ & 1.03 & 176 & 1.97 & 1.67E-03 & $-$1.54E-04 \\
 & 2 & $g^{\rm s}_1$ & 3.15 & 538 & 7.51 & 5.91E-03 &    9.97E-04 \\
 &   & $g^{\rm c}_2$ & 2.24 & 383 & 3.15 & 2.02E-03 & $-$1.01E-04 \\
 &   & $g^{\rm s}_3$ & 1.72 & 294 & 7.11 & 1.02E-03 &    2.38E-04 \\
\enddata
\end{deluxetable}

\clearpage

\clearpage

\clearpage


\clearpage

\clearpage

\clearpage

\end{document}